# Theoretical Investigation of High-$T_c$ Superconductivity in Sr-Doped La$_3$Ni$_2$O$_7$ at Ambient Pressure

Lei Shi[1], Ying Luo[1], Wei Wu[1,2,§], and Yunwei Zhang[1,2,§]

[1]School of Physics, Sun Yat-sen University, 510275 Guangzhou, China
[2]Guangdong Provincial Key Laboratory of Magnetoelectric Physics and Devices, Sun Yat-sen University, Guangzhou 510275, China

**Abstract**

The recent discovery of pressure-induced superconductivity in La$_3$Ni$_2$O$_7$ has established a novel platform for studying unconventional superconductors. However, achieving superconductivity in this system currently requires relatively high pressures. In this study, we propose a chemical pressure strategy via Sr substitution to stabilize high-$T_c$ superconductivity in La$_3$Ni$_2$O$_7$ under ambient conditions. Using density functional theory (DFT) calculations, we systematically investigate the structural and electronic properties of Sr-doped La$_{3-x}$Sr$_x$Ni$_2$O$_7$ (x = 0.25, 0.5, 1) at ambient pressure and identify two dynamically stable phases: La$_{2.5}$Sr$_{0.5}$Ni$_2$O$_7$ and La$_2$SrNi$_2$O$_7$. Our calculations reveal that both phases exhibit metallization of the σ-bonding bands dominated by Ni-$d_{z^2}$ orbitals—a key feature associated with high-$T_c$ superconductivity, as reported in the high-pressure phase of La$_3$Ni$_2$O$_7$. Further analysis using tight-binding models shows that the key hopping parameters in La$_{2.5}$Sr$_{0.5}$Ni$_2$O$_7$ and La$_2$SrNi$_2$O$_7$ closely resemble those of La$_3$Ni$_2$O$_7$ under high pressure, indicating that strong super-exchange interactions between interlayer Ni-$d_{z^2}$ orbitals are preserved. These findings suggest that Sr-doped La$_3$Ni$_2$O$_7$ is a promising candidate for realizing high-$T_c$ superconductivity at ambient pressure.

## 1. Introduction

Nickelates have long been proposed as candidates for cuprates-like unconventional superconductivity due to the electronics similarities between Ni and Cu. However, experimentally observed superconducting critical temperature ($T_c$) in nickelates, such as thin-film Nd$_6$Ni$_5$O$_{12}$, Nd$_{1-x}$Sr$_x$NiO$_2$ and Pr$_{1-x}$Sr$_x$NiO$_2$, remain below the boiling point of liquid (77 K).[1-3] Recently, superconductivity with a $T_c$ of approximately 80 K was discovered in single-crystal La$_3$Ni$_2$O$_7$ at high pressure (14.0 - 43.5 GPa),[4] a finding confirmed by subsequent experimental observations.[5,6] The bilayer Ruddlesden-Popper structure of La$_3$Ni$_2$O$_7$ undergoes a pressure-induced phase transition from the orthorhombic *Amam* phase to *Fmmm* phase, coincident with the emergence of superconductivity. It has been demonstrated that the metallization of Ni-$d_{z^2}$ bonding bands in the high-pressure *Fmmm* phase plays an important role in the emergency of the high-$T_c$ state.[4,7-14] However, the requirement of high pressures severely limits the practical application of La$_3$Ni$_2$O$_7$.

Beyond high-pressure synthesis, chemical doping is a widely used strategy for tuning lattice and electronic properties of materials, often stabilizing superconductivity at reduced pressures. For example, in LaH$_{10}$, Y doping increased $T_c$ above 250 K while reducing the required pressure from 200 GPa to 183 GPa.[15,16] In nickelates, Sr doping has been shown to induce superconductivity in infinite-layer NdNiO$_2$ and PrNiO$_2$ films at ambient pressure by modifying Ni-$d$ orbitals occupancy and adjusting the density of states near the Fermi level ($E_F$)[17-20]. These results suggest that chemical doping could provide a promising approach for realizing high-$T_c$ superconductivity in La$_3$Ni$_2$O$_7$ at lower pressures or even ambient conditions.

Recently, Wang et al. reported pressure-induced superconductivity with a $T_c$ of 40 K in Pr-doped La$_3$Ni$_2$O$_7$.[21,22] Similar to pristine La$_3$Ni$_2$O$_7$, the Ni-$d_{z^2}$ orbitals in Pr-doped La$_3$Ni$_2$O$_7$ exhibit splitting near the $E_F$ at ambient pressure. Upon applying pressure, the Ni-$d_{z^2}$ bonding bands become metalized, suggesting a superconducting mechanism akin to that of pristine La$_3$Ni$_2$O$_7$.[22] However, superconductivity in this system still requires a high pressure exceeding 10 GPa. Several theoretical studies have explored the potential for ambient-pressure-superconductivity in La$_3$Ni$_2$O$_7$ through full substitution of La with rare-earth elements.[23-25] However, these studies revealed that rare-earth bilayer nickelates are unstable in the superconducting *Fmmm* phase of La$_3$Ni$_2$O$_7$ at ambient pressure. Although some stable metallic structures, such as Tb$_3$Ni$_2$O$_7$, have been identified, they exhibit significant structural deviations from the *Fmmm* phase, and their superconducting properties have yet to be experimentally verified.

In this work, we systematically investigate the structural and electronic properties of Sr-doped nickelates La$_{3-x}$Sr$_x$Ni$_2$O$_7$ (x = 0.25, 0.5, 1) at ambient pressure. Among these compositions, we identify two stable phases La$_{2.5}$Sr$_{0.5}$Ni$_2$O$_7$ and La$_2$SrNi$_2$O$_7$, which crystalized in the space group of *Pmma* and *Amam*, respectively. Our density functional theory (DFT) calculations reveal that Sr doping shifts the $E_F$ downward while preserving a band structure similar to that of the superconducting *Fmmm* phase near the $E_F$. Both doped phases exhibit metallization of Ni-$d_z{^2}$ bonding bands, suggesting a superconducting mechanism akin to that of high-pressure La$_3$Ni$_2$O$_7$. Furthermore, our tight-binding model analysis indicates that the key hopping parameters between Ni-3$d$ and O-2$p$ orbitals in La$_{2.5}$Sr$_{0.5}$Ni$_2$O$_7$ and La$_2$SrNi$_2$O$_7$ closely resemble those of the high-pressure pristine phase, preserving strong interlayer super-exchange interactions. These results suggest that Sr-doped La$_3$Ni$_2$O$_7$ is a promising candidate for achieving high-$T_c$ superconductivity at ambient pressure.

## 1. Method

First-principles calculations were performed using DFT as implemented in the Vienna *ab initio* simulation package (VASP).[26,27] The Perdew-Burke-Ernzerhof (PBE)[28] exchange-correlation functional within the generalized gradient approximation (GGA) was employed, with a plane-wave cutoff energy of 600 eV. The Projector-Augmented Wave (PAW) method was used to describe the electron-ion interactions.[29] A 19 × 19 × 5 $k$-points mesh was used for self-consistent calculations. Atomic positions were fully optimized until the forces on each atom were below 0.001 eV/Å. To account for electron correlation effects, a Hubbard $U$ correction of 4 eV was applied to the Ni $3d$ electrons using the GGA+$U$ approach.[30,31] Dynamical stability was examined via density functional perturbation theory (DFPT) as implemented in PHONONPY.[32] Tight-binding Hamiltonians were constructed using the Wannier90 software package based on maximally localized Wannier functions.[33,34]

## 2. Results and Discussion

The crystal structures of La$_{3-x}$Sr$_x$Ni$_2$O$_7$ (x = 0.25, 0.5, 1) at ambient pressure are constructed by substituting La with Sr in the *Amam* phase of pristine La$_3$Ni$_2$O$_7$. In this phase, two distinct La sites exist, corresponding to the Wyckoff position 4c and 8g, labelled as La1 and La2 in Fig. 1(a). La1 atoms occupy the higher-symmetry 4c sites, positioned between the corner-sharing NiO$_6$ octahedral layers, while La2 atoms reside at the lower symmetry 8g sites, each coordinated with eight O atoms. For La$_{2.75}$Sr$_{0.25}$Ni$_2$O$_7$, we identify two polymorphs by substituting a single La atom with Sr at different sites within the unit cell. However, both structures exhibit dynamic instability, as indicated by imaginary frequencies in their phonon spectra (Fig. S1 in the Supplementary Information). For

La$_{2.5}$Sr$_{0.5}$Ni$_2$O$_7$ and La$_2$SrNi$_2$O$_7$, we examined three types of high-symmetry structures, and select the configuration with the lowest-energy for further discussion. Despite the absence of superconductivity observed in recent experiments for Sr-doped La$_3$Ni$_2$O$_7$ with Sr occupying the 8g La sites[35], our calculations indicate that the lowest-energy structures favor for Sr substitution at the 4c position. Detailed structural information and relative energies of all polymorphs for La$_{3-x}$Sr$_x$Ni$_2$O$_7$ (x = 0.5 and 1) are provided in the Supplementary Information (Figs. S2 and S3).

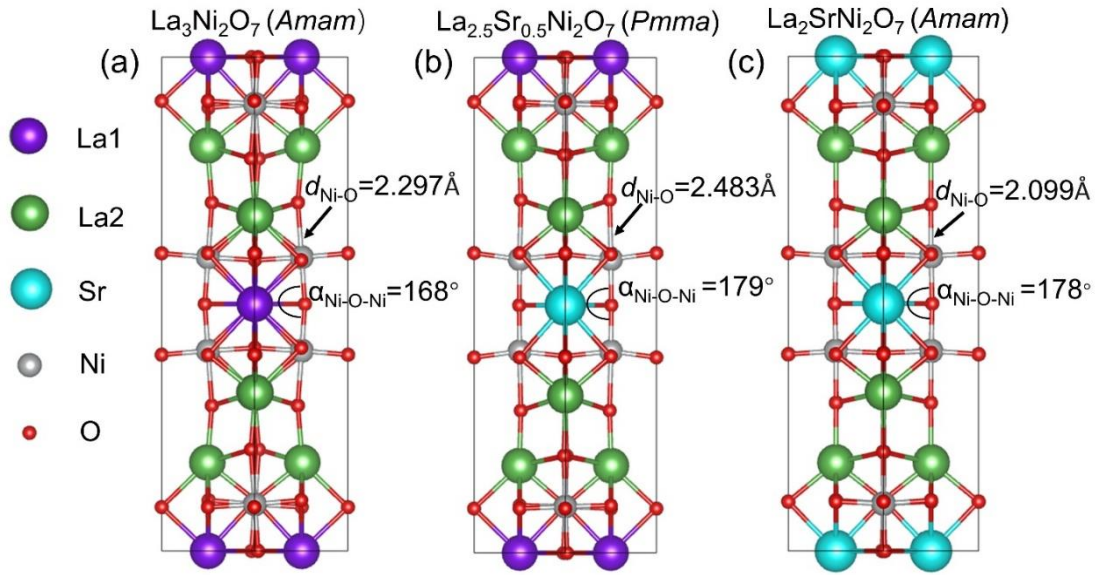

Figure 1. Crystal structures of La$_{3-x}$Sr$_x$Ni$_2$O$_7$ (x = 0, 0.5, 1) at ambient-pressure: (a) *Amam* phase of La$_3$Ni$_2$O$_7$, (b) *Pmma* phase of La$_{2.5}$Sr$_{0.5}$Ni$_2$O$_7$, (c) *Amam* phase of La$_2$SrNi$_2$O$_7$.

La$_{2.75}$Sr$_{0.25}$Ni$_2$O$_7$ adopts a structure with the *Pmma* space group (Fig. 1(b)), while La$_2$SrNi$_2$O$_7$ shares the same structural prototype as the *Amam* phase of La$_3$Ni$_2$O$_7$ (in Fig. 1(c)). Phonon calculations confirm that both structures are dynamically stable (Fig. 2). The corresponding structural information are listed in Table I. In the high-pressure phase of $La_3Ni_2O_7$, the Ni-O (apical oxygen) bond length decreases from 2.297 Å to 2.132 Å, while the Ni-O-Ni bond angle increases from 168° to 180°. For $La_{2.5}Sr_{0.5}Ni_2O_7$ at ambient pressure, the Ni-O bond length is 2.483 Å, slightly larger than that of the $La_3Ni_2O_7$ at 29.5 GPa, due to the larger ionic radius of Sr compared to La. However, the Ni-O-Ni bond angle increases significantly to around 179 Å, indicating that Sr doping induces a chemical pressure effect, which may enhance interlayer coupling. In the case of $La_2SrNi_2O_7$, the Ni-O bond length is 2.099 Å, and the Ni-O-Ni bond angle is 178°, both closely resembling those of the high-pressure phase of $La_3Ni_2O_7$.

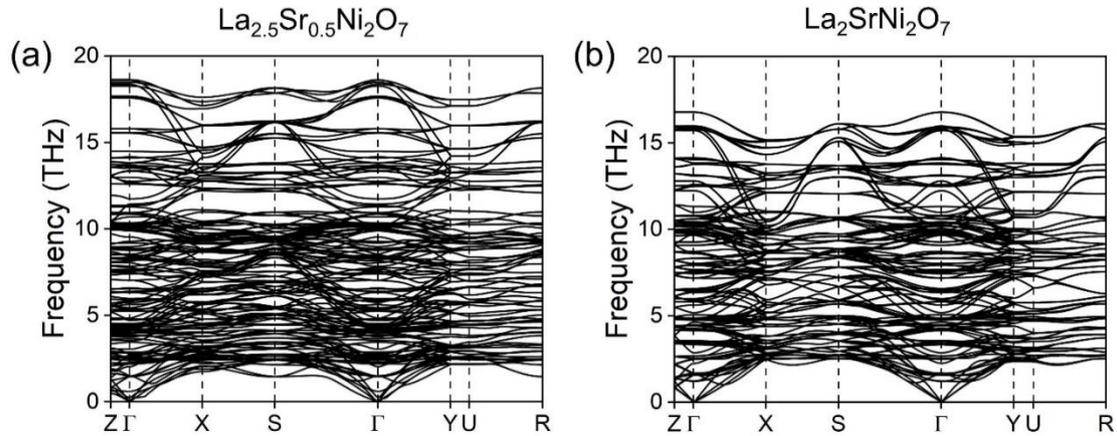

Figure 2. Phonon dispersion of $La_{2.5}Sr_{0.5}Ni_2O_7$ (a) and $La_2SrNi_2O_7$ (b) at ambient pressure.

Table I: Structural information of $La_{3-x}Sr_xNi_2O_7$ (x = 0, 0.5, 1). The pressure, lattice parameters, distance between Ni and optical oxygen ($d_{Ni-O}$), the Ni-O-Ni bonding angle ($\alpha_{Ni-O-Ni}$) are listed.

| Compound | Pressure | a (Å) | b (Å) | c (Å) | $d_{Ni-O}$ | $\alpha_{Ni-O-Ni}$ |
|---|---|---|---|---|---|---|
| La$_3$Ni$_2$O$_7$ | 0 GPa | 5.359 | 5.413 | 20.612 | 2.297 Å | 168° |
| La$_3$Ni$_2$O$_7$ | 29.5 GPa | 5.289 | 5.218 | 19.734 | 2.132 Å | 180° |
| La$_{2.5}$Sr$_{0.5}$Ni$_2$O$_7$ | 0 GPa | 5.367 | 5.359 | 20.753 | 2.483 Å | 179° |
| La$_2$SrNi$_2$O$_7$ | 0 GPa | 5.443 | 5.444 | 20.082 | 2.099 Å | 178° |

We further investigate the electronic properties of these stable phases to explore the potential high-$T_c$ superconductivity in Sr-doped La$_3$Ni$_2$O$_7$. Our band structure calculations for the ambient-pressure *Amam* phase and the superconducting *Fmmm* phase of La$_3$Ni$_2$O$_7$ under high pressure (Figs. 3(a) and 3 (b)) show excellent agreement with previous theoretical studies.[4] The bands associated with Ni-$d_{z^2}$ bonding states become metallized (shown in bule in Fig. 3(b)) at high pressure, which has been identified as a key indicator of superconductivity.[4,7-14] Similarly, our calculated band structures of La$_{2.5}$Sr$_{0.5}$Ni$_2$O$_7$ and La$_2$SrNi$_2$O$_7$ (Figs. 3(c) and 3(d)) at ambient pressure structures also exhibit metallized Ni-$d_{z^2}$ bonding bands. Notably, these band structures near the $E_F$ closely resemble that of the superconducting *Fmmm* phase, which can be attributed to the minimal orbital contribution from Sr atoms near the $E_F$.

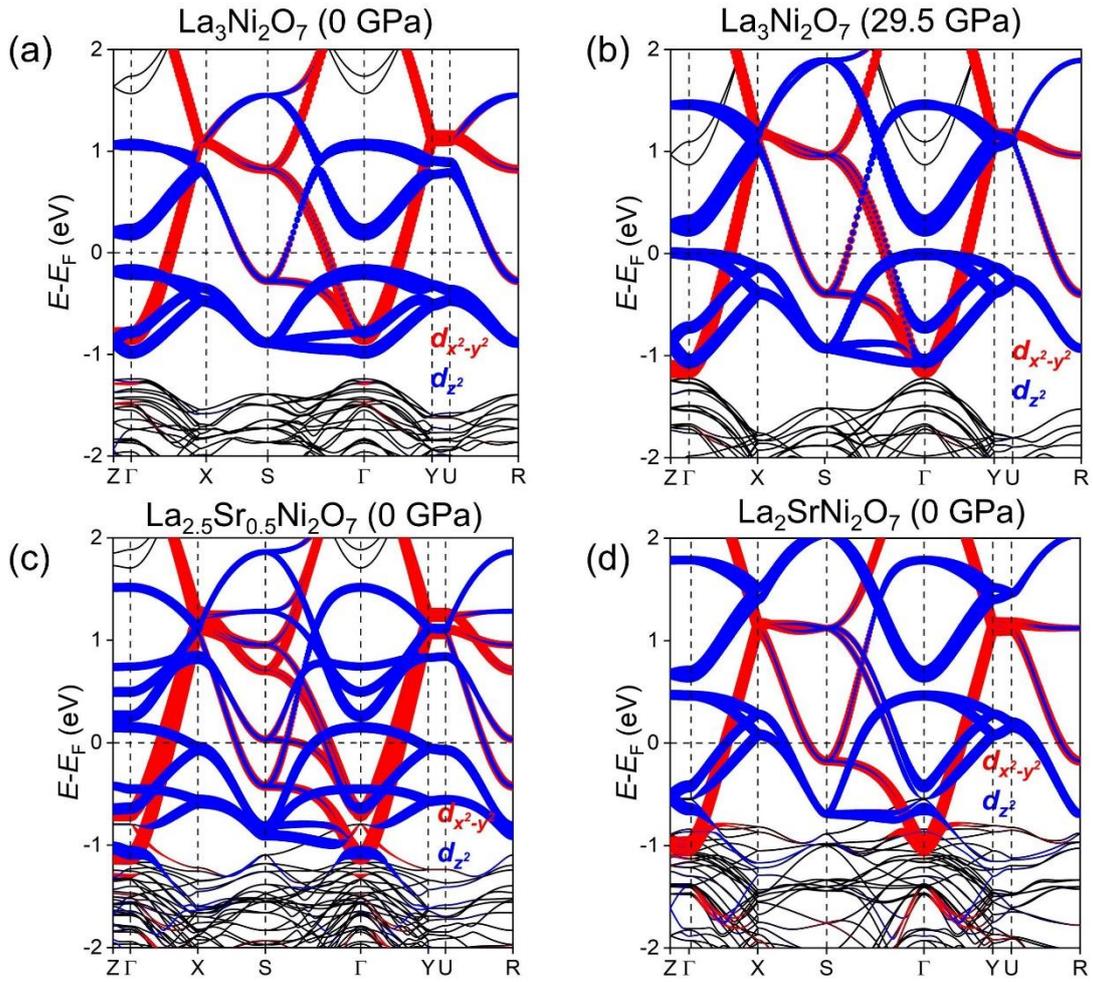

Figure 3. Band structures of $La_3Ni_2O_7$ at 0 GPa (a) and 29. 5 GPa (b), $La_{2.5}Sr_{0.5}Ni_2O_7$ at 0 GPa (c) and $La_2SrNi_2O_7$ at 0 GPa (d). The projections onto Ni-$d_{x^2-y^2}$ and Ni-$d_{z^2}$ orbitals are highlighted in red and blue, respectively. The black dashed lines indicate the Fermi level ($E_F$) in each panel.

As shown in Fig. 3(c) and 3(d), despite the splitting of degenerate orbitals around *S* point and *Z* points in $La_{2.5}Sr_{0.5}Ni_2O_7$ due to the reduced symmetry (Fig. 3(c)), the key band characteristics associated with the superconductivity in the high-pressure phase of $La_3Ni_2O_7$ are preserved in both $La_{2.5}Sr_{0.5}Ni_2O_7$ and $La_2SrNi_2O_7$. The bonding-antibonding splitting of

Ni-$d_{z^2}$ orbitals in La$_{2.5}$Sr$_{0.5}$Ni$_2$O$_7$ and La$_2$SrNi$_2$O$_7$ is approximately 1.4 eV, close to that of 1.5 eV splitting in the high-pressure pristine phase. This suggests that the interlayer hopping amplitude ($t_\perp$) between Ni-$d_{z^2}$ orbitals remains relatively large in these two Sr-doped structures, maintaining strong interlayer super-exchange interactions, which are considered the primary driving force behind superconductivity in high-pressure phase of La$_3$Ni$_2$O$_7$. The electronic interactions between NiO$_2$ bilayers are not significantly weakened at ambient pressure, indicating that La$_{2.5}$Sr$_{0.5}$Ni$_2$O$_7$ and La$_2$SrNi$_2$O$_7$ could host high-$T_c$ superconductivity.

As for the Ni-$d_{x^2-y^2}$ orbitals, the bandwidth in the high-pressure phase of La$_3$Ni$_2$O$_7$ is around 4.2 eV (Fig. 3(b)). In comparison, the Ni-$d_{x^2-y^2}$-derived bands in La$_{2.5}$Sr$_{0.5}$Ni$_2$O$_7$ and La$_2$SrNi$_2$O$_7$ exhibit similar bandwidths at ambient pressure (Fig. 3(c) and Fig. 3(d)), whereas the ambient-pressure phase of La$_3$Ni$_2$O$_7$ has a significantly narrower bandwidth of about 3.6 eV (Fig. 3(a)). This suggests that Sr doping effectively induces a chemical pressure effect, replicating the electronic environment of the high-pressure phase.

Additionally, O-$2p$ orbitals make a significant contribution to the electronic states near the $E_F$ in La$_{2.5}$Sr$_{0.5}$Ni$_2$O$_7$ and La$_2$SrNi$_2$O$_7$ (Figs. S4 (b) and 4(c) in Supplementary Information), strongly hybridizing with both Ni-$d_{z^2}$ and Ni-$d_{x^2-y^2}$ orbitals. This hybridization closely resembles that observed in the high-pressure phase of La$_3$Ni$_2$O$_7$ (Fig. S4 (a) in

Supplementary Information). Thereby the mobilization of spin-singlet pairs and the global phase coherence are maintained in $La_{2.5}Sr_{0.5}Ni_2O_7$ and $La_2SrNi_2O_7$. Meanwhile, Ni-$t_{2g}$ orbitals in $La_{2.5}Sr_{0.5}Ni_2O_7$ and $La_2SrNi_2O_7$ remain far from the $E_F$, suggesting they play a minimal role in superconductivity.

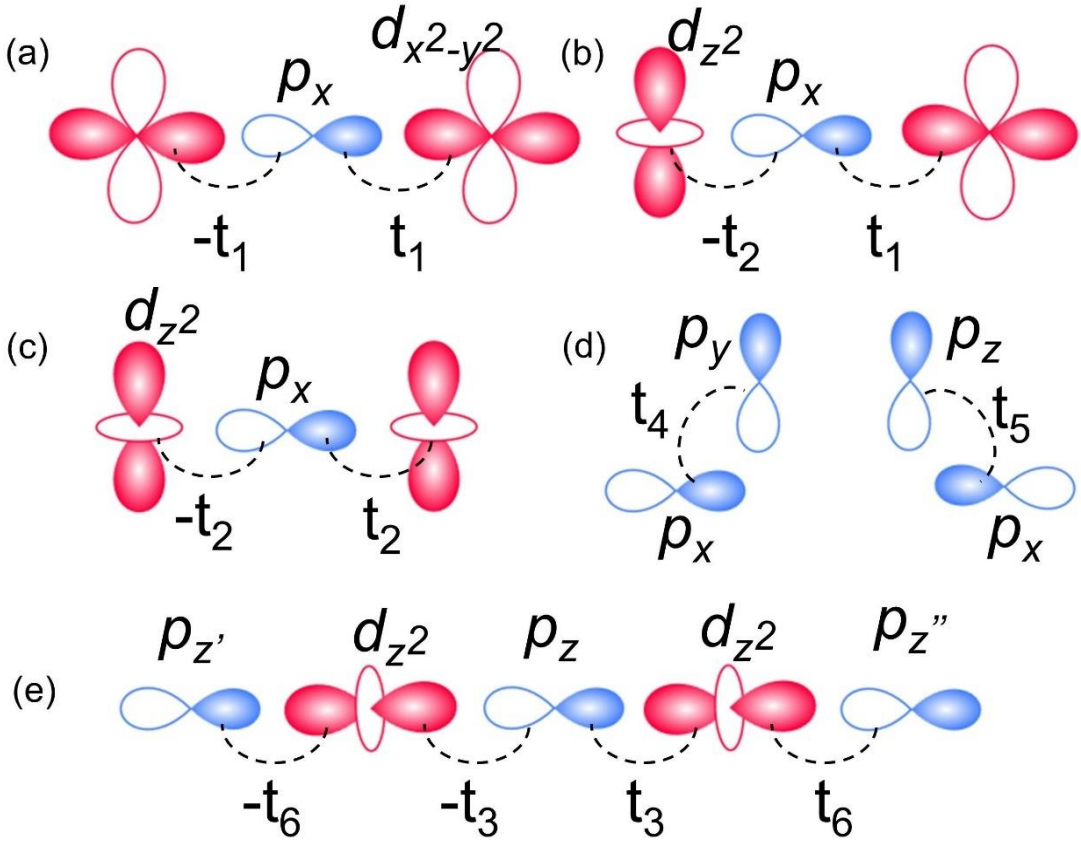

Figure 4. Six hopping processes between Ni-3$d$ and O-2$p$ orbitals, including Ni-$d_{x^2-y^2}$ and O-$p_x$ (a), Ni-$d_{z^2}$ and O-$p_x$, O-$p_x$ and Ni-$d_{x^2-y^2}$ (b), Ni-$d_{z^2}$ and O-$p_x$, (c), O-$p_x$ and O-$p_y$, O-$p_x$ and O-$p_z$ (d), Ni-$d_{z^2}$ and O-$p_z$ (e). The relevant hopping parameters for $La_{3-x}Sr_xNi_2O_7$ (x = 0, 0.5, 1) are listed in Table II.

Table II: The hopping parameters between Ni-3$d$ and O-2$p$ of

$La_{3-x}Sr_xNi_2O_7$ (x = 0, 0.5, 1).

| Compound | Pressure | $t_1$ | $t_2$ | $t_3$ | $t_4$ | $t_5$ | $t_6$ |
|---|---|---|---|---|---|---|---|
| $La_3Ni_2O_7$ | 29.5 GPa | −1.56 | 0.75 | −1.63 | 0.58 | 0.49 | 1.37 |
| $La_{2.5}Sr_{0.5}Ni_2O$ | 0 GPa | −1.39 | 0.64 | −1.54 | 0.59 | 0.40 | 0.73 |
| $La_2SrNi_2O_7$ | 0 GPa | −1.39 | 0.74 | −1.56 | 0.54 | 0.45 | 1.25 |

To further assess the potential for superconductivity in Sr-doped $La_3Ni_2O_7$ at ambient pressure, we construct a tight-binding model for $La_{2.5}Sr_{0.5}Ni_2O_7$ and $La_2SrNi_2O_7$ and compare the relevant hopping parameters with those of $La_3Ni_2O_7$ at high pressure[36,37] (as listed in Table II). For these two Sr-doped structures, the hopping parameter $t_1$, which corresponds to the interaction between Ni-$d_{x^2-y^2}$ and O-$p_x$, is approximately −1.39, slightly smaller than the value in $La_3Ni_2O_7$ at 29.5 GPa ($t_1 = -1.56$). This difference may stem from the larger lattice parameters $a$ and $b$ after Sr doping (Table I). Another key hopping parameter, $t_3$, governs the strength of the super-exchange interactions between interlayer Ni-$d_{z^2}$ orbitals. Notably, $t_3$ remains nearly identical after Sr doping, with values of $t_3 = -1.54$ for $La_{2.5}Sr_{0.5}Ni_2O_7$, $t_3 = -1.56$ for $La_2SrNi_2O_7$, and $t_3 = -1.63$ for $La_3Ni_2O_7$ at 29.5 GPa. This suggests that the effective hopping amplitude $t_\perp$ between interlayer Ni-$d_{z^2}$ orbitals remains relatively large, and that the strong interlayer super-exchange interactions between Ni-$d_{z^2}$ orbitals are maintained in $La_{2.5}Sr_{0.5}Ni_2O_7$ and $La_2SrNi_2O_7$. Furthermore, the other hopping parameters in $La_{2.5}Sr_{0.5}Ni_2O_7$ and $La_2SrNi_2O_7$ closely resemble those in $La_3Ni_2O_7$ at 29.5 GPa. As a result, the energy scales defined by the Fermi energy in the Sr-doped phases align well with those of the superconducting phase at high pressure. This strongly indicates that Sr-doped $La_3Ni_2O_7$ may potentially exhibit high-$T_c$ superconductivity at ambient pressure.

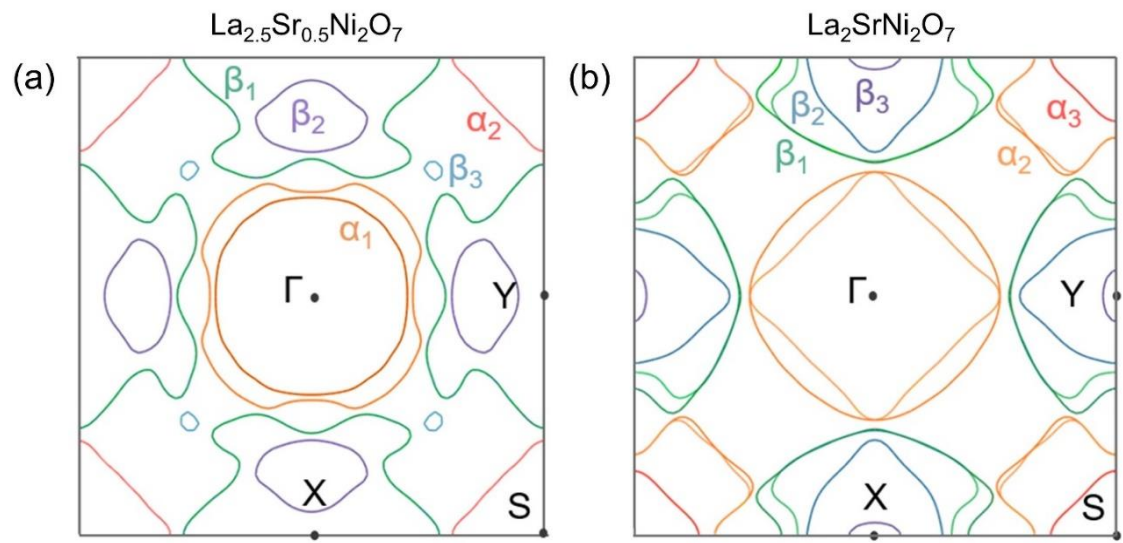

Figure 5. Calculated two-dimensional Fermi surfaces of $La_{2.5}Sr_{0.5}Ni_2O_7$ (a) and $La_2SrNi_2O_7$ (b) within the Brillouin zone at ambient pressure, outlined by a black square. The Fermi surfaces of $La_{2.5}Sr_{0.5}Ni_2O_7$ consist of electrons pockets ($\alpha_{1,2}$) and hole pockets ($\beta_{1,2,3}$). The Fermi surfaces of $La_2SrNi_2O_7$ consist of electrons pockets ($\alpha_{1,2,3}$) and hole pockets ($\beta_{1,2,3}$).

For $La_{2.5}Sr_{0.5}Ni_2O_7$, the $E_F$ shifts downwards by approximately 0.37 eV compared to the pristine $La_3Ni_2O_7$, resulting in the emergence of two new hole pockets ($\beta_2$ and $\beta_3$) along the $\Gamma$ - $X$ and the $\Gamma$ - $S$ paths, respectively

(Fig.5 (a)). However, For the $La_2SrNi_2O_7$, the $E_F$ shifts by 0.639 eV, resulting in a larger electron pocket ($\alpha_2$) around $S$ point and the emergence of two new hole pockets ($\beta_{2,3}$) near the $X$ point. Importantly, these new pockets, primarily composed of Ni-$d_{z^2}$ orbitals, persist at the $E_F$ at ambient pressure and are essential for the emergence of superconductivity. It has been widely discussed that superconductivity in $La_3Ni_2O_7$ follows an $S\pm$ wave pairing symmetry,[8,25,38,39] where spin-singlet pairs form mainly between interlayer Ni-$d_{z^2}$ orbitals within the unit cell, making them localized in the $a$-$b$ plane. Thus, the deformation of the quasi-2D Fermi surface in $La_{2.5}Sr_{0.5}Ni_2O_7$ and $La_2SrNi_2O_7$, compared to the high-pressure $La_3Ni_2O_7$ (Fig. S5), should have minimal impact on superconductivity.

## 3. Conclusion

Our calculations show that Sr doping in $La_3Ni_2O_7$ stabilizes two dynamically stable phases at ambient pressure, $La_{2.5}Sr_{0.5}Ni_2O_7$ and $La_2SrNi_2O_7$, both of which exhibit key features of high-$T_c$ superconductivity. The preserved electronic structures, including the metalized Ni-$d_{z^2}$ bonding bands and strong interlayer super-exchange interactions, suggest that Sr doping effectively mimics the high-pressure environment necessary for superconductivity. This chemical pressure strategy provides a viable route to achieve high-$T_c$ superconductivity in

La$_3$Ni$_2$O$_7$ at ambient conditions, eliminating the need for high-pressure synthesis and enabling practical applications.

ACKNOWLEDGEMENTS

Y.W. Zhang acknowledges funding from the National Key R&D Program of China No. 2023YFA1610000, National Natural Science Foundation of China under Grant No.12304036, the Open Project of Guangdong Provincial Key Laboratory of Magnetoelectric Physics and Devices (No. 2022B1212010008), the Guangdong Basic and Applied Basic Research Foundation (2023A1515010071), and the Fundamental Research Funds for the Central Universities, Sun Yat-sen University (23xkjc016). W. Wu acknowledges funding from the National Natural Science Foundation of China (No. 12494594, No. 12274472).